# Prototype of a bistable polariton field-effect transistor switch

H. Suchomel[1,*], S. Brodbeck[1,*], T.C.H. Liew[2], M. Amthor[1], M. Klaas[1], S. Klembt[1], M. Kamp[1], S. Höfling[1,3], C. Schneider[1]

Microcavity exciton polaritons are promising candidates to build a new generation of highly nonlinear and integrated optoelectronic devices. Such devices range from novel coherent light emitters to reconfigurable potential landscapes for electro-optical polariton-lattice based quantum simulators as well as building blocks of optical logic architectures. Especially for the latter, the strongly interacting nature of the light-matter hybrid particles has been used to facilitate fast and efficient switching of light by light, something which is very hard to achieve with weakly interacting photons. We demonstrate here that polariton transistor switches can be fully integrated in electro-optical schemes by implementing a one-dimensional polariton channel which is operated by an electrical gate rather than by a control laser beam. The operation of the device, which is the polariton equivalent to a field-effect transistor, relies on combining electro-optical potential landscape engineering with local exciton ionization to control the scattering dynamics underneath the gate. We furthermore demonstrate that our device has a region of negative differential resistance and features a completely new way to create bistable behavior.

[1]Technische Physik and Wilhelm-Conrad-Röntgen-Research Center for Complex Material Systems, Universität Würzburg, D-97074 Würzburg, Am Hubland, Germany. [2]Division of Physics and Applied Physics, School of Physical and Mathematical Sciences, Nanyang Technological University, 637371 Singapore, Singapore. [3]SUPA, School of Physics and Astronomy, University of St. Andrews, St. Andrews, KY 16 9SS, United Kingdom. *These authors contributed equally to this work.

Encoding and transporting information by the means of light has been shown to have significant advantages over the use of classical electronic transport. Modern fiber technology allows photons to propagate over large distances (typically several kilometers up to hundreds of kilometers) with hardly any losses and optical devices can, in principle, be modulated faster than electric ones with lower energy dissipation [1]. However, due to their (truly) bosonic nature it is exceedingly hard to make photons interact with each other [2].

Exciton-polaritons (polaritons) - hybrid quasi-particles formed by the strong coupling between quantum well excitons and microcavity photons - can, due to their excitonic matter component, be manipulated easily by external forces. Findings of e.g. Bose-Einstein condensation [3] and superfluidity [4] of polaritons have led to a general description as quantum fluids of light [5]. Polariton fluids can be optically manipulated and interfered [6 - 8] and they can coherently propagate several tens of micrometers in GaAs based planar and channel structures [9 - 11].

Based on these properties, optical circuits on the basis of polaritons have been discussed [12, 13] and all-optical polariton switches [14 - 16] and transistors [17] have been realized. In these implementations, typically a laser creates a local blueshift of the polariton energy due to density-dependent polariton-exciton interactions which offers the possibility to imprint an artificial potential landscape for the propagating polaritons [18].

While steering and switching polaritons with multiple lasers is versatile and reconfigurable, the manipulation via external electric fields is clearly preferable as a significantly more compact and scalable solution. Moreover, the manipulation via external electric fields offers the possibility to investigate fast and reconfigurable manipulation of expanding polariton condensates in channels [8, 10], electro-optical initialization of dark soliton trains [19] and it gives a dynamical tool to engineer fully reconfigurable potential landscapes for a new generation of electro-optical polariton-lattic based quantum simulators [20].

Although initial works have already suggested the feasibility of such an approach [21], it has only been a recent work that has shown a great potential to manipulate the polariton condensate energy on demand in either red or blue shifted direction by the means of the well-known quantum-confined Stark effect (QCSE) on the one hand and due to the controlled reduction of the Rabi splitting on the other [22, 23]. However, while electrically pumped polariton lasers do exist [24, 25], an electrically controlled polariton logic device has not been demonstrated yet. In addition, the influence of an external electric field on polariton propagation is a completely unexplored field and leads, in our case, to the discovery of a completely new way to create bistable behavior.

In this work, we implement an electro-optical polariton transistor switch. The prototype is based on a one-dimensional channel to guide a propagating polariton condensate. We show that a local electrical gate can be used to manipulate the propagation of the polaritons via a combination of the QCSE and locally enhanced polariton dissipation.

**Results**
**Implementation.** Our polariton device is based on a microcavity composed of AlAs/AlGaAs mirrors with 4 GaAs quantum wells (QWs) integrated into the optical antinode of an intrinsic $\lambda/2$-thick AlAs cavity layer. The top mirror is carbon doped (p-type), whereas the bottom mirror is n-type doped using silicon to allow electro-optical tuning of the QWs in the microcavity region. The Rabi splitting amounts to $\hbar\Omega_R = (8.7\pm0.1)$ meV as determined from low temperature white-light reflectance measurements on the unprocessed wafer (see Fig. S1a in the supplementary material). The Q-factor of the microcavity was previously experimentally determined to exceed $Q \sim 8000$.

In order to generate a one-dimensional polariton channel, microwires with a length of 400 µm and a width of 5 µm were defined via optical lithography and etched deeply into the structure using electron-cyclotron-resonance reactive-ion-etching. In a subsequent step, the sample was planarized with the transparent polymer Benzocyclobuthene which serves as a platform for the following contacting steps, but also prevents oxidation of the exposed wire sidewalls. The n-contact on the backside of the substrate is formed by evaporating an AuGe-Ni-Au alloy. Next, the position of the 10 µm wide electrical gate was defined via optical lithography and evaporated a 600 nm thick Au-Ti-layer. A final lift-off step completes the device process (see Fig. 1a).

**Power dependent emission features.** At first the power dependent photoluminescence (PL) emission of our device was investigated via momentum resolved spectroscopy. The sample is mounted in a helium flow cryostat and is oriented parallel to the entrance slit of the spectrometer. A constant helium flow is cooling the sample down to a temperature of ~ 5 K. Polaritons are injected close to the contact by a non-resonant continuous wave laser which is tuned to the reflectivity minimum of the first high-energy Bragg mode around 1.664 eV. The Gaussian shaped pump-spot has a diameter of ~ 5 µm. A longpass filter with a cut off energy of 1.653 eV was placed in front of the spectrometer to filter any scattered light from the laser. Furthermore, a pinhole in the image plane in front of the spectrometer allows us to spatial filter the PL collected from the excited microwire. With an almost completely closed pinhole, we collect luminescence from an approximately 20 µm long region of the microwire. Under moderate power excitation of 2.8 mW and in the closed pinhole configuration a characteristic set of three parabolic dispersions was observed (see Fig. 1b, left) which we identify as the ground state of the microwire as well as higher order lateral modes [10, 26]. The spectra were fitted using a standard coupled oscillator model by coupling the three lowest lateral cavity modes with the heavy hole exciton, yielding a heavy hole exciton-photon detuning of the ground state of δ = -13.6 meV which is slightly larger than the Rabi splitting. We neglected the light hole exciton in this calculation because at this position it lies approximately 50 meV above the cavity energy and therefore should play no crucial role. With increasing pump power, polariton condensation takes place at (4.5 ± 0.5) mW (see supplementary material Fig. S1b), which manifests itself in a massive occupation of a distinct energy state (Fig. 1b). The slight asymmetry of the sample perpendicular to the direction of the wire combined with an imperfect positioning of the pump spot leads to condensation in the second confined subband. This is a result of the open dissipative nature of the condensate, where the gain profile, the modal losses and the intermode relaxation determine the condensate state [27]. A detailed estimation of the corresponding exciton density can be found in the supplementary material. Due to the repulsion from the reservoir at the location of the pump laser, the condensate propagates along the wire direction which can be seen in Fig. 1b. The excitation is located close to the contact that blocks the PL signal underneath, so that the propagation away from the contact ($k_\parallel > 0$ towards red detuned sample gradient) appears more distinct than the one through the contact ($k_\parallel < 0$ towards blue detuned sample gradient) [cf. 28].

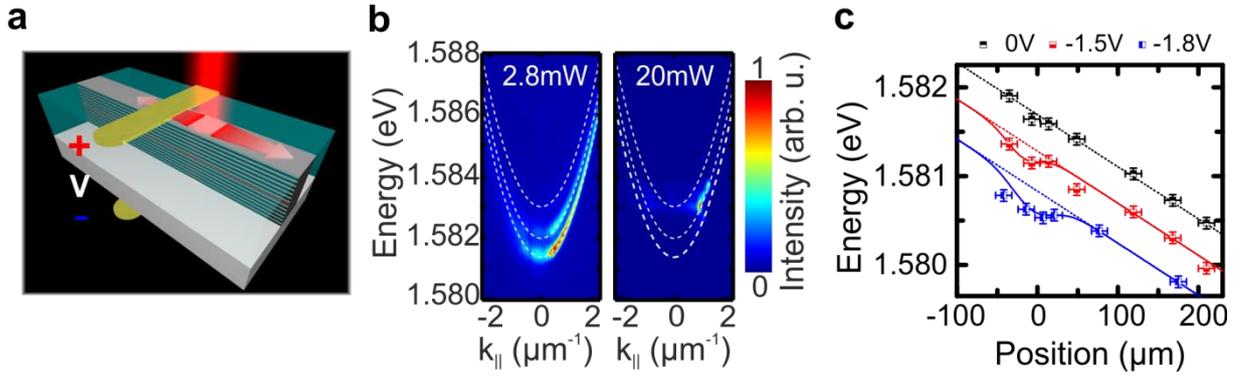

**Figure 1. QW-microcavity polariton transistor switch sample and photoluminescence emission features.** (**a**) Schematic image of the processed device consisting of the microwire and the electrical gate on top of the device. The excitation spot is located close to the contact, generating a propagating polariton flow. (**b**) Momentum resolved photoluminescence measurement at low power (2.8 mW, left) and far above the condensation threshold (20 mW, right) under open circuit conditions. Several lateral modes are visible and fitted by a coupled oscillator model (white dashed lines), yielding an exciton-photon detuning of δ = -13.6 meV for the ground state with respect to the heavy hole exciton. Due to the gold contact, which blocks the PL underneath, the propagation for $k_\parallel > 0$ appears much brighter. (**c**) Spatially resolved PL measurements around zero in-plane wave vector, at low excitation power and several voltages. The reverse bias results in a general redshift of the intrinsic detuning gradient (dotted lines). Furthermore, a local energy minimum evolves underneath the 10 µm wide contact centered at zero position with increasing reverse bias which can be described by an additional Gaussian shaped function (solid line).

**Influence of the electrical gate.** In order to assess the influence of the local gate on the potential landscape along the microcavity wire, we have carried out spatially resolved PL studies under low excitation conditions and in the closed pinhole configuration while changing the applied reverse bias. As can be seen in Fig. 1c, application of a reverse bias results in a general redshift of the ground state emission energy at zero in-plane wave vector which seems to be independent from the position along the 400 µm microwire. However, one can also observe a local minimum of the emission energy underneath the contact (at positions which were not fully covered by gold) which, as we show in the following, will be sufficient to initiate the switching process. The measured spatial extent (full width at half maximum) of the contact-induced potential dip is strongly dependent on the applied voltage. We find a spreading

3/15

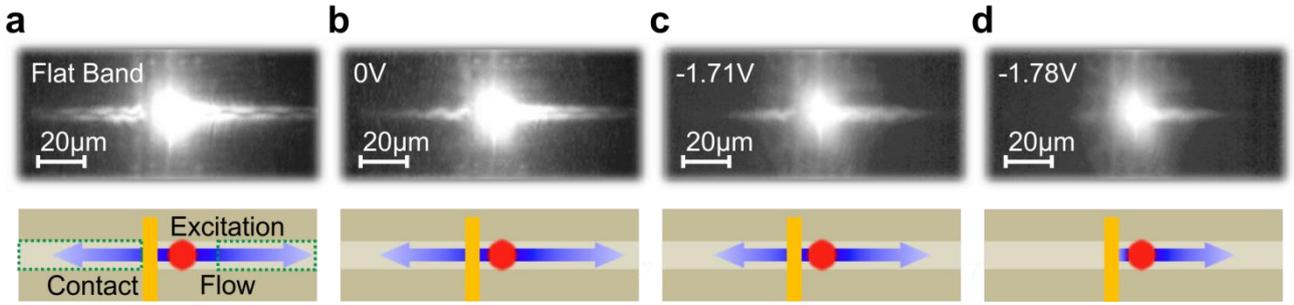

**Figure 2. Polariton switching along the wire.** Real space images of the studied microwire sample under optical excitation with the center of the pump spot approximately 15 µm away from the center of the 10 µm wide electrical contact (top) and schematic drawings of the experimental configuration (bottom). At the contact a voltage is applied and varied between flat band condition ($U_G$ = 1.8 V) and tilted bands ($U_G$ = -1.78 V). Flat band conditions are reached when the external applied bias has equalized the built-in potential in the pin-doped structure and therefore leading to a nearly flat electronic band structure. Furthermore, tilted band specifies voltages in reverse bias leading to an increased built-in potential and therefore to a strongly tilted electronic band structure. The real space images show the flow for different voltages (**a**) $U_G$ = 1.8 V, (**b**) $U_G$ = 0 V, (**c**) $U_G$ = -1.71 V and (**d**) $U_G$ = -1.78 V, respectively, at an excitation power of 20 mW. A clear flow to the right, away from the contact, and to the left, through the contact with respect to the excitation spot, is visible under flat band conditions, while with increasing negative voltage the flow through the contact stops abruptly. The green dashed boxes in the left most schematic drawing mark the integration ranges for the polariton flow through as well as away from the contact, discussed in Fig. 3.

ranging from zero at zero bias to ~ 50 µm at -1.8 V. Data fitting was performed by using an additional Gaussian shaped function superimposed on the intrinsic, linear detuning gradient along the wire due to the layer thickness gradient occurring during sample growth and due to the general red shift. We believe that the doped DBRs work as an additional capacitor able to affect the depletion zone in the pin-doped structure along the whole microwire leading to the general redshift. Typically, structures for indirect excitons in electrostatic potential traps [29, 30] are ni-doped to avoid this problem and to create sharp local potential patterns. This approach is not readily possible in our case since the intrinsic region needs to be as small as possible to obtain large electric fields that shift the exciton energy distinctly. For reverse bias the QCSE leads to a redshift of the emission and results in a potential minimum underneath the contact of 180 µeV and 300 µeV for $U_G$ = -1.5 V and -1.8 V, respectively. The expansion of the condensate along the wire can be directly monitored in an open pinhole configuration by imaging the emission onto a CMOS camera which is mounted in the beam path with a standard optical microscope configuration. The reflection of the excitation laser is filtered out by a longpass filter. Fig. 2 depicts real space images of the studied microwire under optical excitation with the center of the pump spot approximately 15 µm away from the center of the 10 µm wide electrical contact (top). Corresponding schematic drawings of the experimental configuration are shown below the microscope images (bottom). At the contact, a voltage is applied which is varied between flat band condition ($U_G$ = 1.8 V) and strongly tilted bands ($U_G$ = -1.78 V). Flat band conditions are reached when the external applied bias has equalized the built-in potential in the pin-doped structure and therefore leading to a nearly flat electronic band structure. Furthermore, tilted band specifies voltages in reverse bias leading to an increased built-in potential and therefore to a strongly tilted electronic band structure. The real space images show the flow for different voltages $U_G$ = 1.8, 0, -1.71 and -1.78 V, respectively, at an excitation power of 20 mW. Clearly, we can evidence an expansion of the polariton condensate along the wire resulting from the repulsion with the exciton reservoir induced by the pump spot. Under flat band condition a nearly symmetric propagation of the condensate is observed (Fig. 2a). The residual asymmetry in intensity is induced by the already mentioned intrinsic potential gradient in Fig. 1c. It accounts for a soft barrier to the left and thus suppresses the flow in this direction. As we apply a reverse bias, the overall intensity is reduced. More significantly, however, between $U_G$ = -1.71 V and -1.78 V the flow to the left, through the contact, is fully blocked, leading to a strongly asymmetric propagation (Fig. 2c and d). It can also be seen in Fig. 2d that the flow to the right, away from the contact, is significantly reduced after the switching as a result of enhanced losses in the device due to carrier tunneling [31]. The tunneling process corresponds to exciton electric field ionization which therefore affects the polariton decay rate. Evidence for increased exciton electric field ionization with increased negative voltage is given by the bias and spatial dependent exciton lifetime measurements in the supplementary material.

In order to assess the capability of our device to switch the polariton flow in a more quantitative manner, the spatially integrated emission from the first subbranches of the visible wire modes from both sides of the pump spot is plotted as a function of the gate voltage in Fig. 3a. The associated excitation conditions are the same as in Fig. 2. That means the center of the pump spot is located 15 µm away from the center of the 10 µm wide gold contact. The spatially integrated emission of the polariton flow away from the contact is represented by the red triangles, whereas the flow through the contact is plotted as blue circles. The emission is integrated over a range of ~ 40 µm along the length of the wire, starting at a distance approximately 20 µm away from the excitation spot, and over the whole width of the wire (cf. schematic drawings in Fig. 2). With increasing negative voltage the flow in both directions decreases continuously due to increasing tunneling losses. At a voltage of $U_G = -1.60$ V the flow through the contact features a sharp intensity drop, which we identify as the threshold from the "on"-state to the "off"-state, while the flow away from the contact shows a more gentle drop at around $U_G = -1.72$ V. As our numerical calculations in the discussion below show, the flow is blocked by a soft potential barrier for $U_G < -1.60$ V and fully trapped for $U_G < -1.72$V. Furthermore, the gentler drop of the propagating polariton condensate away from the contact is due to the smaller influence of the enhanced carrier losses underneath the contact (see Fig. S2 in the supplementary material). The integration range for the spatially integrated emission away as well as though the contact start at the edge of the potential dip evolving under the contact. Therefore, a drop of the integrated emission is seen in both cases, first when the polariton flow gets blocked at the transition from the "on"-state to the "off"-state and second when the condensate gets fully trapped. The exact behavior of the trapped polariton condensate cannot be tested directly in this experimental configuration since most of the potential well is covered by the deposited gold contact. However, in our case the spatially dependent polariton decay rate, being enhanced underneath the contact, has to be taken into account and the expected increase of the polariton emission for the trapped polariton condensate [16] is not obvious. Considering this, our numerical calculations in the discussion below suggest only a small variation of the trapped polariton emission close to the excitation spot. In order to further visualize the pronounced asymmetry between the spatially integrated emission of the polariton flow away and through the contact, the corresponding ratio is plotted in Fig. 3b. Switching between the 'on-' and 'off-state' by the local gate is unambiguously verified in this representation.

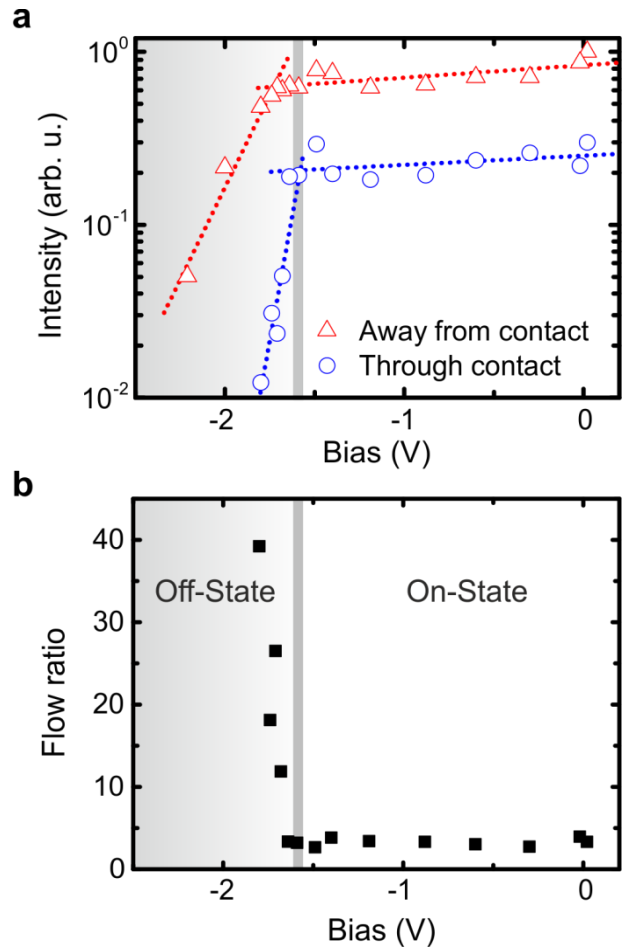

**Figure 3. Quantitative analysis of the switching process.** (**a**) Spatially integrated emission of the propagating polariton condensate away from the contact (red triangles) and through the contact (blue circle) for an excitation power of 20 mW. The center of the excitation spot is approximately 15 µm away from the center of the 10 µm wide electrical contact. Dotted lines are guides to the eye. The emission is integrated over a range of ~ 40 µm along the length of the wire, starting at a distance approximately 20 µm away from the excitation spot, and over the whole width of the wire (cf. schematic drawings in Fig. 2). With increasing negative voltage the flow in both directions decreases continuously. At a voltage of $U_G$ = -1.60 V the flow through the contact shows a sharp intensity drop, while the flow away from the contact shows a slighter drop at around $U_G$ = -1.72 V. (**b**) Corresponding ratio between the flow away and through the contact. A sharp increase of the ratio is visible at around $U_G$ = -1.64 V, indicating the 'off-state' of the flow direction.

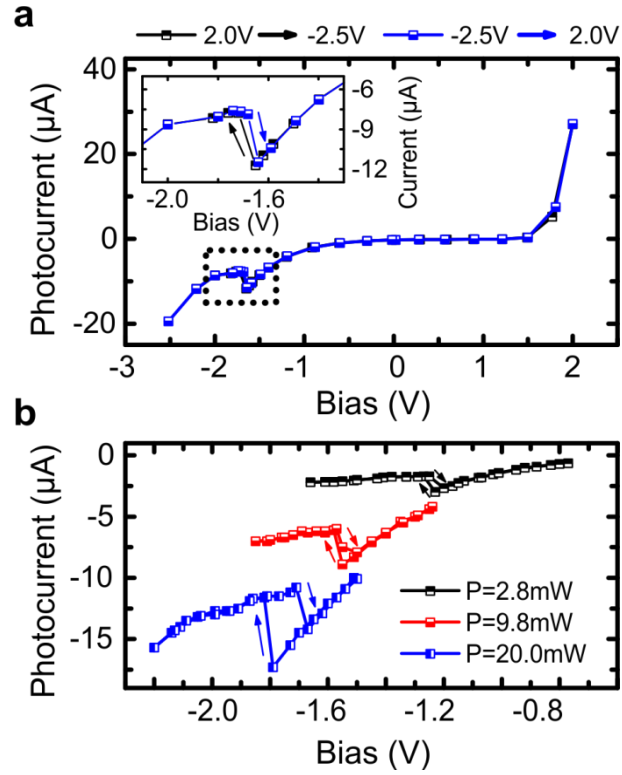

**Figure 4. Photocurrent readout and bistable switching.** (**a**) Measured photocurrent in forward as well as in reverse direction for an excitation power of 20 mW, with the pump spot approximately 25 µm away from the center of the contact. The black and blue squares indicate the varied voltage from 2.0 V to -2.5 V as well as from -2.5 V to 2.0 V, respectively. Between -1.62 V and -1.72 V a negative differential resistance is visible in the photocurrent corresponding to the on-/off-state of the device. The inset shows a zoom into the region of negative differential resistance. (**b**) By moving the pump spot closer to the contact, which means approximately to a distance of 15 µm to the center of the contact, the nonlinearity shows a bistable behavior with upward and downward directed bias. By varying the pump power and therefore the injected carrier density the bistability emerges at 20 mW or vanishes for lower excitation powers.

We can furthermore exploit the electrical contact as a sensitive probe of the carriers which diffuse and propagate along the wire. Therefore, we extracted the photocurrent for various excitation powers by measuring the voltage drop across a series resistance with R = 1.0 kΩ. In Fig 4a, we kept the excitation power at 20 mW and located the pump spot approximately 20 µm away from the contact. While the device features the standard behavior of a diode, between $U_G$ = -1.62 V and -1.72 V a marked negative differential resistance is visible which appears at approximately the same applied negative voltage as the threshold where the device switches between the "off"- and the "on"-state. The inset shows a zoom into the region of negative differential resistance. By moving the pump spot closer to the contact, which means to a distance of approximately 10 µm, a strong bistable behavior evolves as we increase the negative voltage in the device. As we show in Fig. 4b, this hysteresis is a sensitive function of the pump power, and as such, of the injected carrier density.

**Discussion**

We use a simplified 1D model based on the incoherently driven Gross-Pitaevskii approach [32, 33] to model our findings. Coherent polaritons are described by the order parameter, ψ(x,t), the dynamics of which is given by:

$$i\hbar \frac{d\psi(x,t)}{dt} = \left[ -\frac{\hbar^2 \hat{\nabla}^2}{2m} + (\alpha - i\alpha_{NL})|\psi(x,t)|^2 - \frac{i\Gamma(x)}{2} + V(x) + iP(x) \right] \psi(x,t)$$

where m is the polariton effective mass, α is the strength of polariton-polariton interactions and $\alpha_{NL}$ is the strength of nonlinear losses. The polariton decay rate Γ(x) is spatially dependent, being enhanced underneath the contact where the applied electric field induces exciton dissociation. The losses in the system are balanced by the continuous wave pumping P(x), to form steady states in the system. The effective potential of polaritons V(x) includes the experimentally measured linear potential gradient β. Furthermore we consider a repulsive potential peak due to the hot excitons excited at the pump position, characterized by constant g, and an additional potential dip at the position of the electrical contact $V_C(x)$, which can be varied in strength:

V(x) = βx + gP(x) + $V_C$(x)

The spatial distribution of the additional potential dip $V_C(x)$ is chosen in agreement with the experimental values extracted from Fig. 1c.

We solved the driven-dissipative Gross-Pitaevskii equation numerically for a slowly varying contact-induced potential depth $V_C$, where for each value of $V_C$ the system formed a quasi-stationary state. The polariton intensity spatially integrated beyond the contact (i.e., the total intensity of polaritons that passed the contact) and the polariton intensity spatially integrated over the contact position are shown in Fig. 5a and b as functions of $V_C$, respectively. We note that we can observe a switching behavior at a potential depth of ~ $V_C$ = 0.75 meV. At this point, the polariton flow through the contact is blocked, and the device acts as a switch, analogous to our experimental findings in Fig. 3. Stopping the polariton flow through the contact simultaneously acts on the population underneath the contact which is plotted in Fig. 5b. In fact, this population features the bistable behavior which we have previously probed via the photocurrent. To provide a better understanding of these effects, we plot the expanding polariton cloud for various applied electric fields in Fig. 5c-e. For flat-band condition and small bias, repulsion of polaritons from the incoherent exciton reservoir is the dominating effect [34], leading to a nearly symmetric real space pattern (Fig. 5c). In agreement with the experimental observed propagation at flat band conditions (cf. Fig 1c), the linear potential gradient works as a soft barrier and thus suppresses the propagation. Beyond a threshold (~ 0.75 meV), the effect of the applied electric field becomes dominant and reconfigures the potential landscape in such a way that only propagation away from the contact is possible (Fig. 5d). Larger contact-induced potential depths lead to a full trapping of condensed polaritons (Fig. 5e). Besides the "on"- and the "off"-state this third "trapped"-state regime can also be identified in the experimental data of Fig. 3a where the integrated emitted intensity of the polariton flow away from the contact starts to drop. It is in principal possible for the system of polaritons to demonstrate bistability due to a switching between condensed modes inside the contact-induced trap [35]. Such a bistability is strongly pronounced in the dependence of the polariton intensity integrated over the contact as a function of increasing or decreasing $V_C$. The two different stationary states that appear in the bistable region have slightly different energies and spatial structure. Mechanisms of bistable switching based on resonant excitation are well-known [36 - 38] in exciton-polariton systems and recently a variety of works have reported non-resonantly excited bistability, both theoretically [39, 40] and experimentally [41, 42] in particular configurations. However, these mechanisms do not rely on the spatial degree of freedom of polaritons.

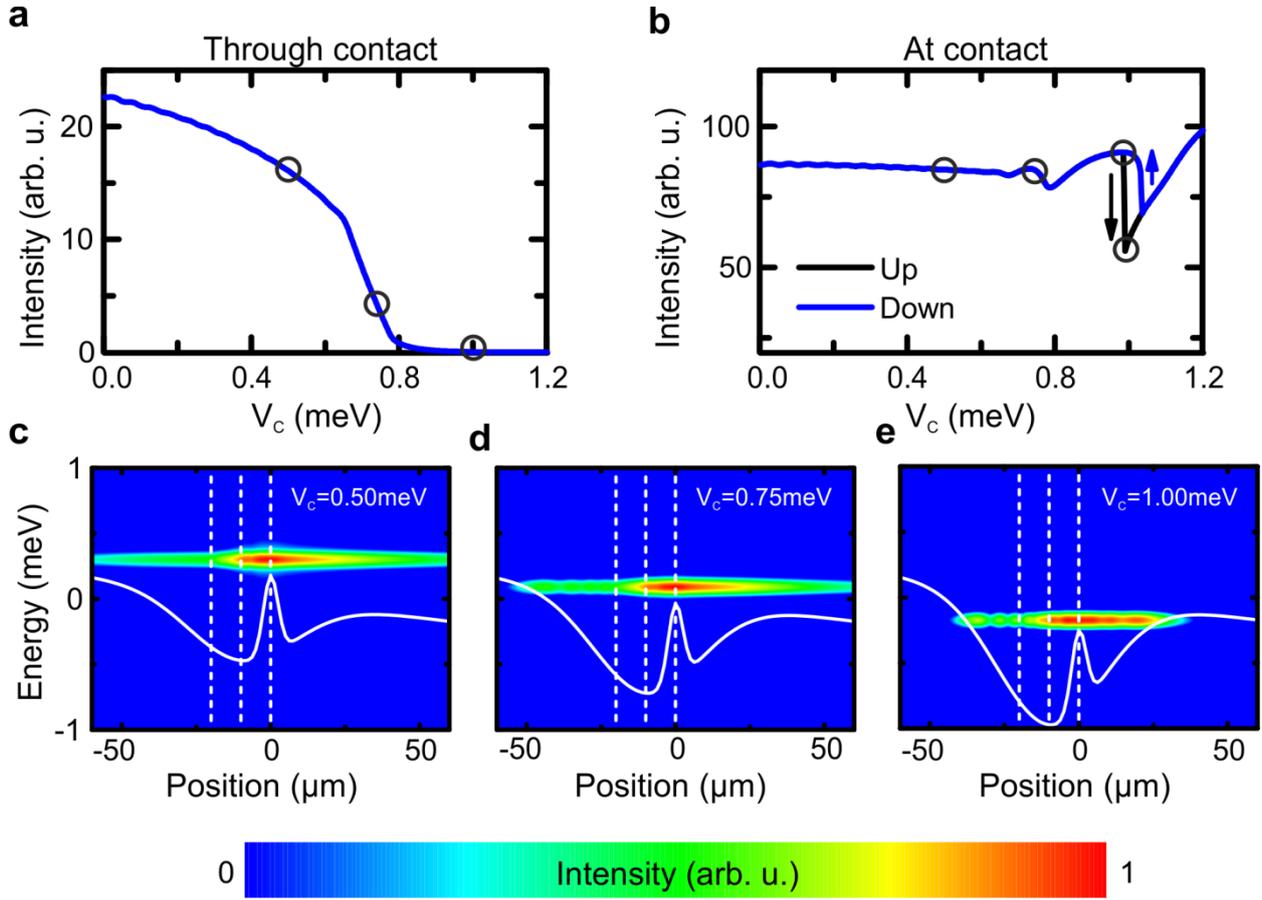

**Figure 5. Gross-Pitaevskii model of the experiment.** (**a-b**) Total intensity of polaritons that passed the contact (**a**) and before the contact (**b**). (**c-e**) Potential landscape for the injected polariton flow around the contact for different contact-induced potential depths, $V_C$. Without an applied potential dip there is a propagation of polaritons through the contact and away from it (**c**). At a potential depth of ~ 0.75 meV only the polariton flow through the contact is blocked (asymmetric case) (**d**) while the propagation is finally trapped at higher values of $V_C$ (**e**).

The bistability in our device of light appears due to a transition between polariton condensation modes in the presence of the contact-induced trap. The potential trap supports confined polariton modes, which compete with each other as they are fed by the non-resonant pump that injects polaritons on the edge of the trap. The mode favored in the competition is typically the one with largest overlap with the non-resonant pump. As the electric potential is changed, the depth of the potential trap also changes, which allows changes in the mode with greatest overlap with the non-resonant pump. The bistability results near a potential depth where there is a change in the favored confined mode. This is because if a particular mode has been previously populated it experiences enhanced gain due to stimulated scattering processes. This introduces a dependence on the history of the system, where a mode that would normally not win the mode competition can do so if it was previously populated under different conditions. The change in the polariton density affects the (stimulated) relaxation rate of hot carriers and excitons in the system. Therefore the appearance and thus the width of the bistability region is determined on how long the enhanced gain due to stimulated scattering processes compensates the lack of the reduced mode overlap with the non-resonant pump. Accordingly the width of the bistability region is a sensitive function of the carrier reservoir. The corresponding change in its density is directly reflected in the experimentally observed photocurrent response (see Fig.4b).

Besides the fundamental prospects, namely the trapping of a propagating polariton condensate in an electrostatic potential as well as the bistable behavior as a probe for a fundamental gain process, polariton devices offer some advantages compared to conventional silicon based transistor technologies. First, the group velocity, that means the velocity to transport information, as well as switching times had been shown to be at least one order of magnitude higher [16, 17]. Furthermore, activation energies required for the switching process are significantly small (~ 1 fJ) [17]. Anyhow, for the construction of complex polariton networks the system needs to satisfy several qualitative criteria [1]. The most important criteria are cascadability, logic-level restauration, logic level independent of system losses and amplification. Those criteria are discussed in detail in the supplementary material. We believe that by taking advantage of the well-controlled bistability and with the combination of existing work, one can satisfy all of those criteria. However, we suggest applications for a single device where one needs to control a propagating polariton condensate, for example in optical intercore communication. In the long term, further work is needed to show if electro-optical polariton devices in complex polariton networks are capable to reach a technological market. Major challenges are still the compact realization of several coupled devices and the ability to reach large distances as discussed in the supplementary material.

In conclusion, we have implemented a prototype of an electro-optical polariton transistor switch. The functionality of the device, which is the polariton equivalent of a field-effect transistor in its basic definition, is facilitated via local electro-optical potential shaping, giving a dynamic tool to investigate the influence of a static electric field on polariton propagation. We believe that our work represents an important step towards the implementation of compact and densely packed logic gates based on light-matter coupled hybrid particles. As we have already demonstrated electro-optical readout of the population underneath the gate, we suggest implementing similar schemes to probe the transmitted polaritons as a fully electrical readout. In combination with an electrically injected polariton pump [24, 25], electro-optical interferometers and amplifiers, a full architecture of all-electrical integrated polaritonics is now within reach. In addition to the switching operation, transitions between trapped condensate modes lead to a pronounced negative differential resistance in the photo-current response and a strong bistability which represents a completely new way to create bistable behavior. Furthermore, the accomplished full trapping of a polariton condensate in an electro-static potential yields a flexible tool to engineer fully reconfigurable potential landscapes for polaritons.

**Methods**
**Sample design.** The bottom (top) distributed Bragg reflector (DBR) consists of 27 (23.5) AlAs/ Al$_{0.2}$Ga$_{0.8}$As mirror pairs surrounding a $\lambda/2$-AlAs cavity with a nominal length of 100 nm. A single stack with four 7 nm wide GaAs-QWs is placed in the field maximum inside the cavity, whereat every QW is separated by 4 nm AlAs barriers. To lower the series resistance of the DBRs, all abrupt AlAs/Al$_{0.2}$Ga$_{0.8}$As heterointerfaces have been replaced by 20 nm wide quasi-graded superlattices. Furthermore the bottom (top) mirror is n-doped with silicon (p-doped with carbon) at a doping concentration of $3 \cdot 10^{18}$ cm$^{-3}$. The doping is gradually reduced to $1 \cdot 10^{18}$ cm$^{-3}$ towards the intrinsic cavity. The topmost mirror pairs in the top DBR are highly C-doped at a concentration of $2 \cdot 10^{19}$ cm$^{-3}$ to get a sufficient ohmic contact with an evaporated Au-layer. The whole structure has been grown by molecular beam epitaxy on a Si-doped (001) GaAs substrate.
**Experimental setup.** A compact low temperature PL setup was constructed, in which both spatially (near-field) and momentum-space (far-field) resolved spectroscopy and imaging are accessible. Photoluminescence was collected through a 0.4 NA microscope objective, and directed into an imaging spectrometer with 1200 groves/mm grating via a set of lenses, projecting the proper projection plane onto the monochromator's entrance slit. The momentum space coverage in this configuration amounts to $k_\parallel = \pm 2.2$ µm$^{-1}$ with a resolution of ~ 0.05 µm$^{-1}$ (~ 0.5°). The spectral resolution is ~ 0.05 meV. A nitrogen–cooled Si charge-coupled device was used as detector.
**Gross-Pitaevskii model.** In solving the Gross-Pitaevskii equation, we took the following parameters: polariton lifetime (away from contact), $\hbar/\Gamma = 10$ ps; polariton-polariton interaction strength, $\alpha = 2.4 \times 10^{-3}$ meV µm$^2$; nonlinear loss, $\alpha_{NL} = 0.3\alpha$ [31]. The effective mass, m, was fitted to the low power dispersion as $2.93 \times 10^{-5}$ of the free electron mass. The contact-induced potential was taken with Gaussian width 25 µm, in agreement with low power spatially resolved experimental spectra. The potential gradient was taken as 2.86 eV/m. The parameter g was chosen such that gP(0)=0.6 meV, where P(0)=0.33 meV µm$^{-2}$. The final bistable behaviour was sensitive to a fitting parameter representing the enhanced decay rate beneath the contact, which was allowed to reach 5Γ.

## Acknowledgements

This work has been supported by the State of Bavaria. The authors thank A. Wolf, M. Wagenbrenner and S. Kreutzer for experimental and technical support. C.S. thanks E. A. Ostrovskaya, for fruitful discussions.

**Correspondence and requests** for materials should be addressed to Christian Schneider (christian.schneider@physik.uni-wuerzburg.de)

# Supplementary Information

In the following, additional experimental data and details on the demonstration of a bistable polariton field-effect transistor switch will be provided.

**Optical characterization**

Initially the planar, unprocessed microcavity sample has been characterized via low temperature white light reflectance measurements. By utilizing the intrinsic layer thickness gradient along the radial direction of the wafer the cavity mode can be tuned through the exciton energy resonance. Fig. S1a depicts the energies of the absorption dips from the reflectance spectra, which we relate to the lower (LP) middle (MP) and upper (UP) polariton branches, due to the coupling with the heavy and light hole exciton modes. Both the LP and the MP show a clear anti-crossing with the respective higher polariton branch which is the fingerprint for strong coupling between the cavity mode and the QW excitons. The inset shows the reflectance spectrum around resonance between the LP and MP branch. From the spectra, we extract a Rabi-splitting of $(8.7 \pm 0.1)$ meV.

The etched microwire sample has been characterized under open circuit conditions by investigating the power dependent emission spectra in a momentum resolved spectroscopy setup. Fig. S1b depicts the emission intensity as well as the evolution of the linewidth at the relevant in-plane wave vector where condensation takes place around $\sim k_{\parallel} = 1$ µm$^{-1}$ as a function of the pump power. Data have been acquired on the processed sample, approximately 50 µm away from the contact. From the non-linearity, we can extract a threshold power for polariton lasing of $(4.5 \pm 0.5)$ mW. Furthermore, Fig. S1c depicts the power dependent emission energy of the relevant in-plane wave vector where condensation takes place around $\sim k_{\parallel} = 1$ µm$^{-1}$. The continuous blue shift above the threshold is a clear indicator that strong coupling conditions persist in our processed wire up to 20 mW.

Power dependent photoluminescence was investigated via momentum resolved spectroscopy. Using a standard coupled oscillator model by coupling the three lowest lateral cavity modes with the heavy hole exciton to fit the spectra gives a heavy hole exciton-photon detuning of the ground state of $\delta = -13.6$ meV. From this calculation one gets the Hopfield coefficients for the exciton and the three lateral cavity photon fractions in the ground state of the system. At zero in-plane wave vector the exciton fraction amounts to approximately 8%. The quite large negative exciton-photon detuning at open circuit conditions was chosen because an applied electric field leads to a renormalization of the polariton energy levels [1]. Therefore, with increasing electric field strength the exciton-photon detuning shifts towards smaller absolute values and thereby the exciton fraction in the lower polariton branch increases. Moreover, the propagating polariton condensate carries an in-plane wave vector unequal zero which leads additionally to an increased exciton fraction compared to the value at zero in-plane wave vector. According to that the expected polariton lifetime is a complex function of field strength, position on the wire and pump power leading to an estimated polariton lifetime between 3 and 10 ps.

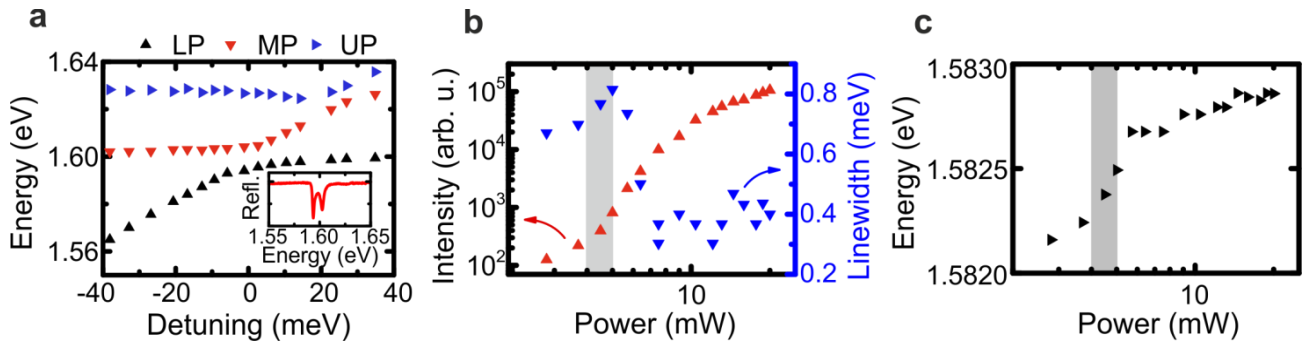

**Figure S1 | Optical characterization for strong coupling and polariton condensation.** (**a**) Energy position of the absorption dips as a function of exciton-photon detuning extracted from low temperature white light reflectance spectra on the planar, unprocessed wafer. The respective absorption dips can be related to the lower (LP), middle (MP) and upper (UP) polariton branches. The inset

shows a section of the reflectance spectrum near resonance between the photon mode and the heavy hole QW-exciton. The Rabi-splitting between the LP and the MP branch determines to (8.7 ± 0.1) meV. (**b**) Power dependent emission intensity (red) and linewidth evolution (blue). (**c**) Power dependent emission energy of the relevant in-plane wave vector where condensation takes place around ~ $k_\parallel$ = 1 µm$^{-1}$. Typical "S"-shape of the intensity accompanied by a sharp linewidth drop at a threshold power of (4.5 ± 0.5) mW in (b) in combination with the continuous blue shift in (c) are clear signatures for polariton condensation.

**Exciton Density Estimate**

We assume a coupling efficiency of the microscope objective of $T_{Obj}$ = 0.8, a transmission of the cryostat window of $T_{Cryo}$ = 0.9, and furthermore a sample reflectivity of $R_{Sample}$ = 0.30 at the pumping wavelength extracted from low temperature reflectivity measurements. According to [2] one can assume 1% absorption per QW. With the given spot size of $d_{Spot}$ = 5 µm and the excitation power of P = 20 mW at a laser energy of $E_{Laser}$ = 1.664 eV one can calculate the exciton density per QW $n_{Ex}$ via:

$$n_{Ex} = \frac{P * T_{Obj} * T_{Cryo} * (1 - R_{Sample}) * A_{QW}}{\pi \left(\frac{d_{Spot}}{2}\right)^2} * \frac{\tau}{E_{Laser}}$$

According to [3] the relevant timescale is the average relaxation time of the excitons scattering into the polariton states. Therefore τ is in the range of 20 ps for the used GaAs QWs [4]. Thereby we get an exciton density of 3.9*10$^{10}$ cm$^{-2}$ which is approximately one order of magnitude smaller than the expected Mott transition at 3*10$^{11}$ cm$^{-2}$ in a GaAs QW system [5]. One can also estimate the polariton density via the observed blueshift of ~ 1 meV and the polariton-polariton interaction strength $\alpha \approx 6E_B a_B^2$ by using standard values for the exciton binding energy $E_B$ and the exciton Bohr radius $a_B$ [6]. Here we get a polariton density of ~ 2*10$^{10}$ cm$^{-2}$.

**Time-resolved measurement**

The spatial dependence of the polariton decay rate Γ(x) along the wire, especially close to the electric gate, can be accessed via time resolved spectroscopy. We excited the microwire at a various positions in the closed pinhole configuration non-resonantly in the first high-energy Bragg mode with a pulsed Ti:sapphire laser. The laser facilitates 50 ps long pulses with a repetition rate of 82 MHz. The corresponding PL signal has been filtered for the exciton line around 1.5957 eV, using the grating position of our spectrometer, and was measured with an avalanche photodiode (APD), connected to the side port of the spectrometer. The spectral bandwidth is given by the grating and the slit size of the side port and amounts to 150 µeV at the given energy. The APD provides a nominal time resolution of 40 ps. Since the exciton lifetime is associated with the polariton lifetime via the Hopfield coefficients and the quality factor, it is a good indication to measure the bias and spatial dependency of the exciton lifetime along the microwire sample to get an insight into Γ(x). The exciton lifetime has been obtained by measuring the actual exponential decay of the exciton reservoir that is built up in the duration of the 50 ps pulse.

Fig. S2a and b depict selected time-resolved measurements of the exciton decay at a distance of 7 µm away from the contact for different applied voltages (a) and at a constant bias of -1.8 V at different positions on the wire (b). In (a) one can see a clear evidence for a faster exciton decay which means a shorter lifetime with an increasing applied reverse bias. However, in (b) one can see a prominent spatial dependence of the exciton decay along the wire. Exponential fitting of the decay slope yields the exciton lifetime which is shown in Fig. S2c as a function of the position along the microwire for several applied voltages. The center of the 10 µm wide contact is located around 0 µm and its width is indicated by the vertical dotted lines. The increasing reverse bias results in a decreasing exciton lifetime and furthermore in a formation of a lifetime minimum right underneath the contact due to enhanced exciton ionization, introduced by the applied electric field. Just like the potential minimum underneath the contact, the lifetime minimum can be described by a Gaussian shaped function indicated by the solid lines. Note that the electric field leading to the exciton ionization typically splits up in a component perpendicular and parallel to the quantum well plane [7]. Both components lead to exciton field ionization but on different field strength scales [8]. Due to the pin-doping the in-plane component is suppressed and the component perpendicular to the quantum well plane should be the most relevant electric field component for the tunneling process.

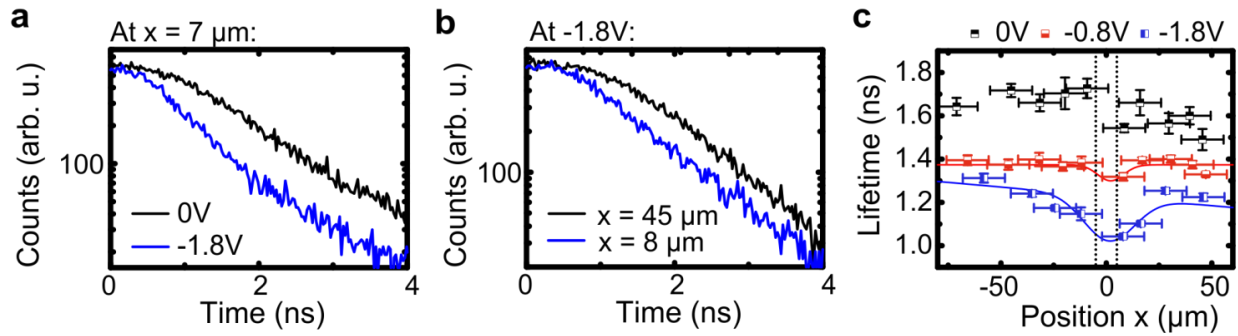

**Figure S2 | Bias and spatial dependence of the exciton lifetime along the wire.** (**a**) Time-resolved measurement of the exciton decay at a distance of 7 µm away from the contact for different applied voltages (0 V black; -1.8 V blue) in semi-logarithmic scale. With increasing reverse bias the excitons decay faster which leads to a shorter lifetime. (**b**) Time-resolved measurement of the exciton decay at a constant applied bias of -1.8 V at different positions along the wire in semi-logarithmic scale. The decay indicates a prominent spatial dependence of the exciton lifetime along the wire. (**c**) Exponential fitting of the respective decay slopes yields the spatially resolved exciton lifetime along the microwire sample for different applied voltages. An increasing reverse bias leads to a general reduction of the exciton lifetime and furthermore to the formation of a lifetime minimum underneath the contact. The minimum can be fitted by a Gaussian shape function (solid lines) and serves as a guide to the eye.

**Cascadability**

To satisfy cascadability, the output of one device must be able to control the output of a second one. For that purpose we extended the incoherently driven Gross-Pitaevskii approach to the case of two devices. While keeping the intrinsic potential gradient one can speak of the upper (1) and the lower switch (2) (cf. Fig. S3b). First of all, both devices can be independently switched between the two stable states of the bistable regime by the application of electric pulses. Remarkably, for electric pulses smaller than the width of the hysteresis region, the state of the lower switch (2) can only be switched if the upper switch (1) is in a specific state. Fig. S3a depicts the intensity at the contact corresponding to the lower switch (2) and upper switch (1). Initially both devices are in the high intensity state. At 2 ns a voltage pulse switches the upper switch into the lower intensity state. At 4 ns, a weak voltage pulse applied to the lower switch (2) is unable to switch its state. At around 8 ns the upper switch (1) is reset to the high intensity state. Due to its coupling to the lower switch this allows the lower switch to be switched when a weak pulse (same amplitude as the pulse at 4 ns) is applied at around 10 ns. This represents the electrical analogue of the cascadability shown by Ballarini et al. [9].

**Logic-level restoration**

The present system is ideal for logic-level restoration since there is a well-controlled bistability. The bistability is triggered by the carrier reservoir in the presence of the contact-induced trap. The bistable behavior vanishes as soon as the pump spot is moved away from the gate. Switching the device between the "off"- to the "on"- state is already sufficient to encode binary information. However, operating in the bistable regime means that the state of any device is necessarily in one of two stable distinguishable states which reduce the impact of noise on the system. Therefore, the quality of the logic signal should be restored at each stage when extending the experiment to the case of several devices and signal degradations should not propagate through the system.

**Logic level independence of system losses**

Working in the bistable regime allows encoding of binary information, where the two bistable states represent clearly distinguishable logic levels. Considering a system made with many devices, it is reasonable to assume that each switch would be defined by an identical hysteresis curve. Hence all switches throughout a circuit would be defined with the same logic-levels.

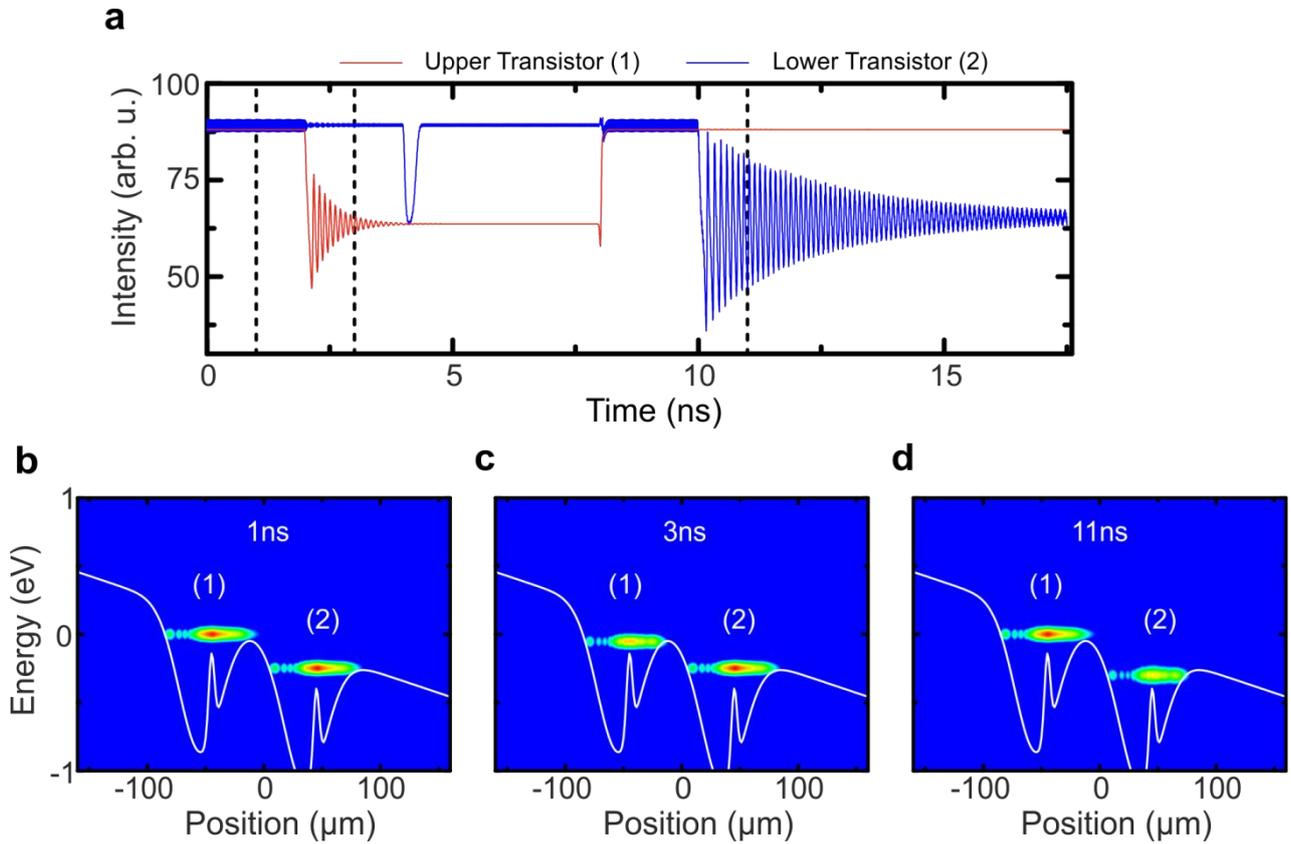

**Figure S3 | Extended Gross-Pitaevskii model for cascadability.** (**a**) Total integrated intensity at the contact of the corresponding upper (1) and lower switch (2). For weak voltage pulses, the state of the lower switch (2) can only be switched when the upper switch (1) is in a specific state. (**b-d**) Potential landscape for the injected polariton flow for the case of two devices at different times corresponding to the time evolution of (**a**). (**b**) At 1 ns both devices are switched to the on-state. (**c**) At 3 ns only the lower switch (2) is switched to the on-state while the upper switch is switched to the off-state. (**d**) At 11 ns the upper switch (1) is switched to the on-state and therefore offers the possibility to switch the lower switch (2) to the off-state.

The calculation shown in Fig. S3 demonstrate that it is possible for the state of one switch to affect logically the state of its neighbor, when the devices are operated in the bistable regime. Within this scheme there should be no loss of information as it propagates through a larger circuit. The reason is that although there are losses of polaritons as they propagate, they are fully compensated by the gain, coming from the incoherent driving of each switch region, when they arrive at the next device. Consequently, there is no need to define logic-levels based on information loss.

**Amplification**

Amplification of a polariton signals in one dimensional channels has already been demonstrated several times [10, 11] in configurations that can be implemented straight forwardly. While in the work of Wertz et al. and Niemietz et al. the amplification of the propagating polariton condensate has been accomplished by providing gain via an optically induced exciton reservoir, it is reasonable to assume that local electrical injection would accomplish the same task. Beyond that, for the case of two devices (see discussion of "cascadability") we believe that amplification is already implemented since losses of polaritons are fully compensated by the gain, coming from the incoherent driving of an additional laser at each switch region.